# Design and development of opto-neural processors for simulation of neural networks trained in image detection for potential implementation in hybrid robotics.

A thesis submitted to the Graduate School of the University of Cincinnati
In partial fulfillment of the requirements for the degree of

**MASTER OF SCIENCE (M.S.)**
**NON-THESIS**

In the Department of Electrical Engineering and Computer Science of the College of Engineering and Applied Science

September 2023

By

**SANJANA PUSHPARAJ SHETTY**

B.Eng. Electronics & Communications Engineering
Visvesvaraya Technological University, 2015

Committee Chair: Dr. Rashmi Jha

# Abstract

Neural networks have been employed for a wide range of processing applications like image processing, motor control, object detection and many others. Living neural networks offer advantages of lower power consumption, faster processing, and biological realism. Optogenetics offers high spatial and temporal control over biological neurons and presents potential in training live neural networks. This work proposes a simulated living neural network trained indirectly by backpropagating STDP based algorithms using precision activation by optogenetics achieving accuracy comparable to traditional neural network training algorithms.

## Acknowledgements

I would like to take this opportunity to thank the most important person because of whom I was able to be part of the laboratory, Dr Rashmi Jha, who provided me the platform to pursue my passion of conducting research on Neurons and Machine Learning applications. I express my sincere gratitude to Dr Jha for her insightful guidance, constant encouragement, and personal help.

I would take this opportunity to express my gratitude to my colleagues and mentors in the MIND Laboratory: Vishal, Abhijeet, Thomas for providing me the help, support, and joyful discussions. I am thankful to the University of Cincinnati, where I gained the necessary background which laid the foundation to my research.

Last but most importantly, I am in sincere debt of gratitude to my parents for constant support, sacrifice, love and care in making me the woman I am today. It wouldn't have been possible without their selfless help and guidance in life.

# Table of Contents

# List of Figures

# Chapter 1: Introduction

Deep neural networks have found extensive applications in every area of the modern world. Extensive research has been done to optimize neural networks to their maximum effectiveness and lowered energy costs. Various attempts have been made to increase the efficiency of a deep learning networks. Biological neural networks have long been considered highly computationally and energy efficient. With the average energy used by the human brain, estimated around 15 kilocalories per hour which is comparable to what a quad core CPU uses in 20 minutes. Biological neurons also can fire up to a maximum of 200 times per second making them faster than solid state electronics. Further, their microscopic sizes allow for a large number of neurons to fit into smaller chips. These reasons are motivation to explore the culturing and training of live neural networks for a variety of applications. Live neural networks require precise stimulation and indirect weight change mechanisms. These requirements are addressed in this work by the use of optogenetic stimulation and development of a novel algorithm for indirect external training.

While precise stimulation and training of neurons is challenging, recent advances in the field of optogenetics, cell-level micropatterning and membrane targeted genetic sensors for neural activity indicators, has made it possible for the required spatial temporal accuracy required to be achieved.

Optogenetics is the encoding of light sensitive opsins in neurons for selective stimulation or inhibition of neuronal activity. The advantages of control, precision and reversibility provide a wide range of possible applications for optogenetic methods. Many groups are currently researching and developing devices for optical stimulation applicable for in vitro and in vivo optogenetics. The major advantage offered by optogenetics is its temporal precision and noninvasive control of activity. Boyden et al. [3] demonstrated that photoactivation of cation permeable Channelrhodopsin2 (ChR2) can regulate the membrane potential and spiking in genetically targeted neurons on a millisecond time scale. A wide range of models exist to simulate and predict the behavior of living neuronal networks. The Hodgkin-Huxley model is a conductance-based model with a high level of bio-realism and real-time operation. Additionally, calcium dependent channel models have been embedded in these realizations for neuronal adaptation. Using these state models in combination, neural processors for simulation and prediction of opto-neural behavior have been developed [8], which can be used to stimulate biologically plausible models.

Neurons in biological networks communicate via spike patterns. The strength of synapse between neurons determines spiking patterns and model function. Biological neural networks offer several constraints that make traditional methods of ANN training impossible to implement. Existing training algorithms depend on direct manipulation of synaptic weights, which is not possible in biological networks. However, synaptic weights can be controlled by forced external stimulation of neurons to trigger spike timing dependent plasticity (STDP) based weight changes. Other constraints involve using spike rates in place of decimal outputs, restriction to only positive conductance values and unalterable activation functions.

In this paper we propose an indirect STDP based backpropagation algorithm used to train biologically plausible neural networks with optogenetic precision.

### 1.1 Traditional Methods of training neuronal cultures

Neural cultures have historically been used as models for the study of underlying learning memory, information processing and network properties. Compared to intracellular neural experiments, neural cultures offer a large range of advantages. Neural cultures offer the option to record from the same set of neurons over a span of several months. These studies can be used to study both structural and functional plasticity. Neural cultures also offer the ease of imaging the neural activity. Multi-electrode arrays (MEAs) are extensively used in these studies. Multi-electrode arrays are extracellular electrodes embedded in cultures. They are used to provide electrical stimuli to neural cultures and readout electrical activity. They have been extensively used to study neural cultures [34].

MEAs though functionally very beneficial for electrical stimulation and readout of neural cultures, possess a handful of drawbacks. Electrical stimulation is less precise in terms of spatial and temporal resolution. Electrical stimulation also makes simultaneous excitation and readout very difficult. Optical stimulation has been proposed to overcome the limitations. Optical stimulation offers precision and a distinguishable external excitation which facilitates single cell excitation of neurons and lends itself as the means needed for indirect training of neural networks mimicking the direct training of artificial neural networks.

### 1.2 Optogenetics

Using optical systems and genetic engineering technologies, optogenetics is a sophisticated method that allows for precise manipulation and observation of the biological activities of a cell, collection of cells, tissues, or organs with exceptional temporal and spatial accuracy. Optogenetics involves: (i) Manipulating a gene to develop reactivity to light, (ii) Delivery of concentrated light to manipulated gene and (iii) studying the effect of the excitation mechanism.

The technique utilizes opsins or light activated proteins to make cells “reactive to light”. Opsins are light-sensitive proteins found in microbes, and act as channels or pumps to facilitate the passage of ions across cell membranes upon activation by light.

### 1.3 Opsins

Opsins are divided into two superfamilies: microbial opsins (type 1) and animal opsins (type 2). These microbial opsins make cells light sensitive i.e the direct stimulation of cells with specific light wavelengths electrically activate neurons. Three types of opsins have been studied over decades, specifically for neural activation:

• Channelrhodopsin: It is activated by blue light of 460 nm wavelength and produces an excitatory effect. Upon electrical transduction induced by light, Channelrhodopsin alters in structure causing an influx of Sodium ($Na^+$) , potassium ($K^+$), Hydrogen ($H^+$) and calcium ($Ca^+$) causing a depolarization of the cell. This depolarization causes the membrane potential to reach its threshold and creates an action potential, effectively causing the neuron to fire. It was discovered in unicellular green algae

• Halorhodopsin: It is activated by yellow light of 570 nm wavelength and produces an inhibitory effect. When activated by light, the protein alters its structure causing an influx of chloride ions ($Cl^-$). This causes the cell to hyperpolarize and inactivate the neuron.

• Bacteriorhodopsin: It is found in mammalian retinas. It pumps out protons (H+) when excited by light.

### 1.4 Methods of delivery

There are three methods to express optogenetic tools into the method:

- Injection of viral vectors: A viral vector is genetically modified and delivered to neurons by targeted genetic identity or circuit connectivity patterns. One of the most common delivery vectors is the adeno-associated virus (AAV). A combination of the opsin, a marker and cell specific promoter is packaged into the viral vector and used to modify neurons.
- Transgenic modification of animals: Transgenesis works instead by breeding an animal that carries the opsin (or any other gene) in its DNA and enhancing them by crossing them with promoter lines.

- In utero electroporation: In this method, a target filled patch pipette is directed to the site of interest using two photon microscopy. A solution with the DNA is inserted into the brains of embryos.

## 1.5 Delivery of Light

Light activated opsins need specific wavelengths of lights for activation. As discussed in the previous section, Channelrhodopsin is activated by ~460nm light and Halorhodopsin is activated by ~570 nm light. The precision of various opsins allows for the coexpression of various opsins together in neural cultures to create combinations of excitatory and inhibitory neurons.

- Laser light sources: Lasers are commonly used in the optogenetic applications due to the high precision (narrow bandwidth) properties and their ability to be linked to optical fibers. This facilitates their use in deep brain stimulation using an implanted fiber optic device. Fiber optics give us flexibility to obtain specific length and structure for devices used in optogenetic stimulation.
- LED light sources: light emitting diode light sources offer advantages such as precise adjustment of spectral sensitivity and low costs. Their minimal power demands and compact size offer advantages in multisite illumination and portable deep brain stimulation devices [32,33]

The biggest advantage optical stimulation offers over electrical stimulation is its high specificity and resolution which provides the ability to perform single cell excitation, thus aiding in precise training of neural cultures. Various groups have developed LED microarrays that can facilitate this level of precision. Streude et all [22] have developed a microarray connected to a high-definition multimedia driver with a flexible connector. Each pixel of the array can be turned on and off by the driver and the CMOS backplane, thus providing controlled light exposure of individual cells. Light-induced changes in cell membrane current are measured with a patch clamp electrode (voltage clamp mode, whole-cell configuration). Braekan et al [24] have developed a multi-electrode-optrode array chip. The light is introduced into the system at the input array region, which is located at the edge of the chip. There is a grating coupler at each location (indicated with a triangle) that couples light into a particular waveguide when a light source is placed above it. The light is then carried by the corresponding waveguide into one of the optrodes (output grating coupler). The MEOA is composed of an array of eight-by-eight optrodes and TiN electrodes. There are two types of optrodes for two different wavelengths (450 nm, corresponding to blue, and 590 nm, corresponding to amber), which are vertically interlaced. The algorithm proposed in this work relies on this type of specificity for its implementation.

## 1.6 Current readout

There are a wide number of readout methods for neural activity. Neural readout devices rely on the ability of integrated circuit technology to capture electrical signals from neurons. Over the years, recording methods have evolved from single cell recordings to development of microelectrode arrays for single cell recordings of larger networks. The basic equipment needed to record neural activity is micromanipulators, amplifiers, microelectrodes and recording devices. The most direct method of recording neural activity is using implantable microelectrodes. Simultaneous recording of presynaptic and postsynaptic recordings has been vital to understanding neural mechanisms of communication. Methods used vary from paired patch clamp recordings of multiple neurons, optical probing technique and calcium-based imaging technique.

# Chapter 2: Spiking Neural Networks

Spiking neural network models are considered to more closely mimic biological neural networks since they are inspired by information processing in biology. SNNs are used to model the brain's enormous parallel processing and communication using sparse and asynchronous binary signals. Spiking neural networks were originally studied as models of biological information processing [31], where neurons exchange information via spikes. SNNs process information using two factors: timing of spikes and identity of synapses (excitatory vs inhibitory neurons). SNNs incorporate time-based information transmission between neurons and also exhibit "bursting" and "synchrony", that are characteristics of biological neural networks. Neurons within the network often communicate with each other through action potential spikes, which typically exhibit a consistent time course. SNNs so closely mimic spiking neural networks in functionality, that they are compatible with sparsity and temporal coding. SNNs also have low power consumption and event driven information processing, thus used in this work to model all the advantages a live neural network offers.

Biological neurons process information by the propagation of electrochemical signals via action potentials. The constant transfer of ions, inside and out of a neuron, causes charges to transfer along the length of a neuron and to other neurons via synapses. Three primary ions that are responsible for electrically charging the neurons are $Na^+$, $K^+$, $Ca^+$ and $Cl^-$. Synapses are responsible for transfer of information in networks. They can be divided into excitatory and inhibitory synapses. Excitatory synapses depolarize postsynaptic neurons and boost the firing of action potentials. Inhibitory synapses hyperpolarize postsynaptic neurons and inhibit the development of action potentials.

There are several neuron models used to model precise biological behavior such as integrate and fire, Izhikevich and Hodgkin- Huxley models.

## 2.1Network model

To propose an algorithm potentially implementable on a biologically realistic network, a simple network with known architecture is used. To exploit the computational advantages offered by live neural networks, networks can be cultured with predefined geometry. A number of methods exist to culture neural networks in a fixed architecture. The methods range from the utilization of microfluidic channels to cellular lithography [9][10]. Numerous studies have combined MEAs with neural culture setups to obtain precisely aligned networks for easy readout.

The method of utilizing microfluidic channels for neural culture neurons employs a process of using hippocampal neurons placed in cell chambers of PDMS chips with adhesion promoting substances. The chips are maintained at optimal conditions and closely monitored to promote cellular growth. The neural cultures adapt to the structure of the microfluidic channel, thus creating the required predefined neural architecture.

The method of cellular lithography has used various techniques such as the use of hydrophilic materials to manipulate neuronal cell attachments, UV techniques, microcontact printing, micro-machined surfaces, microfluidic deposition, photoresist patterning and photoablation. All of these techniques are used to promote neural growth in required architectures.

**2.2 Proposed Network model**

In this work, the network proposed is a three layer fully connected network potentially implementable on a biologically realistic network, a simple network with known architecture is used. To exploit the computational advantages offered by live neural networks, networks can be cultured with, with an equal number of inhibitory and excitatory neurons. A combination of one inhibitory and one excitatory neuron is modeled to replicate a single artificial neuron as biological conductance can only be positive. The number of neurons was varied for different datasets and for optimum accuracy.

**2.3 Neuron model**

Adhering to the requirement of biological plausibility, a neural unit of highest biological realism is implemented. A model of a CA3 hippocampal pyramidal neuron [14] with an integrated CHr2 channel is used to model an optogenetically active neuron. The structure consists of four compartments: Synapses, Axon, Dendrites and Soma. The model has incorporated calcium-dependent ion channels to further replicate the role of Ca ions in synaptic plasticity. Voltage gated calcium ion channels are responsible for the intake of calcium ions caused by depolarization. The resultant neural model can be activated by electrical and light impulses. An alpha synapse was used to account for the input from multiple neurons over a certain period producing a continuous function.

The total membrane change in membrane voltage of a single optogenetically active neuron is given below:

$$CdVmembdt = -\,(Isyn + INa + IKdr + IKa + IKahp + IKc + ICa + IChR2 + ILeak + (gc * Vmemb) \qquad (1)$$

C = Membrane capacitance
Vmemb – Membrane voltage
Isyn – Synaptic current
INa – Sodium ion channel current

IKdr – delayed rectifier potassium ion channel current IKa – A-type of transient potassium ion channel current

IKahp – long duration Ca-dependent $K^+$ ion channel current

IKc – short duration Ca-dependent $K^+$ ion channel current
ICa – Calcium ion channel current
IChR2 – light dependent ion channel current ILeak – Leakage current
gc – conductance

Excitatory synapse:

$$I_{synExcite} = gsyn(t)(V(t) - E_{syn}) \qquad (2)$$

Inhibitory synapse:

$$I_{synInhib} = \quad gsyn(t)(V(t) + E_{syn}) \qquad (3)$$

Total conductance over time-period 1 to n:

$$gsyn(t) = \sum^{n} gsyn \frac{t-ti}{\tau} exp(\frac{-t-ti}{\tau}) \quad i{=}1 \qquad (4)$$

**2.4 STDP and Calcium**

The amplitude of the postsynaptic calcium ion concentration determines the direction of potentiation or depression in excitatory neurons [35]. Two key steps have been recognized as responsible for the generation of synaptic plasticity in hippocampus cells. NMDA receptors and voltage dependent $Ca^{2+}$ govern the entry of postsynaptic calcium. This influx of calcium then triggers LTD (Long term depression) and LTP (Long term potentiation) in post-synaptic hippocampal neurons. Traditionally, high frequency synaptic stimulation, which suggests high concentration of $Ca^{2+}$ induces LTP and low frequency synaptic stimulation, which suggests low concentration of $Ca^{2+}$ induces LTD [36]. Spike timing dependent plasticity is a type of Hebbian synaptic plasticity that features simultaneous post and pre-synaptic plasticity with temporal specificity. Studies have shown that the levels of influx of $Ca^{2+}$ are directly related to the strength of synaptic connections [17].

### 2.5 Learning Rule

Artificial neural networks generate precise floating-point numbers as outputs and their training is conducted by direct adjustment of the synaptic weights to calculated weights. Biological networks communicate via all-or-nothing spike trains. Their synaptic weights are indirectly altered by STDP based spiking and must be manipulated to obtain required changes in weights. Single increments/decrements employ Calcium based STDP rules for learning in the presented algorithm. Synaptic weights are changed according to spike times of Presynaptic and Postsynaptic neurons. Equation (5) depicts the learning rule.

$$\frac{dW_i}{dt} = \eta([Ca])(\Omega([Ca]) - \lambda W\mathrm{i} \qquad (5)$$

Where *Wi* represents the synaptic strength of synapse i, η is the learning rate, and the calcium level at synapse is denoted by [Ca]. λWi is a weight decay term that helps stabilize synaptic growth without imposing a saturation limit. dWi then represents the change in synaptic weight in a single potentiation/depression step.

The ϴd and ϴp are selected arbitrarily as a window for depression.

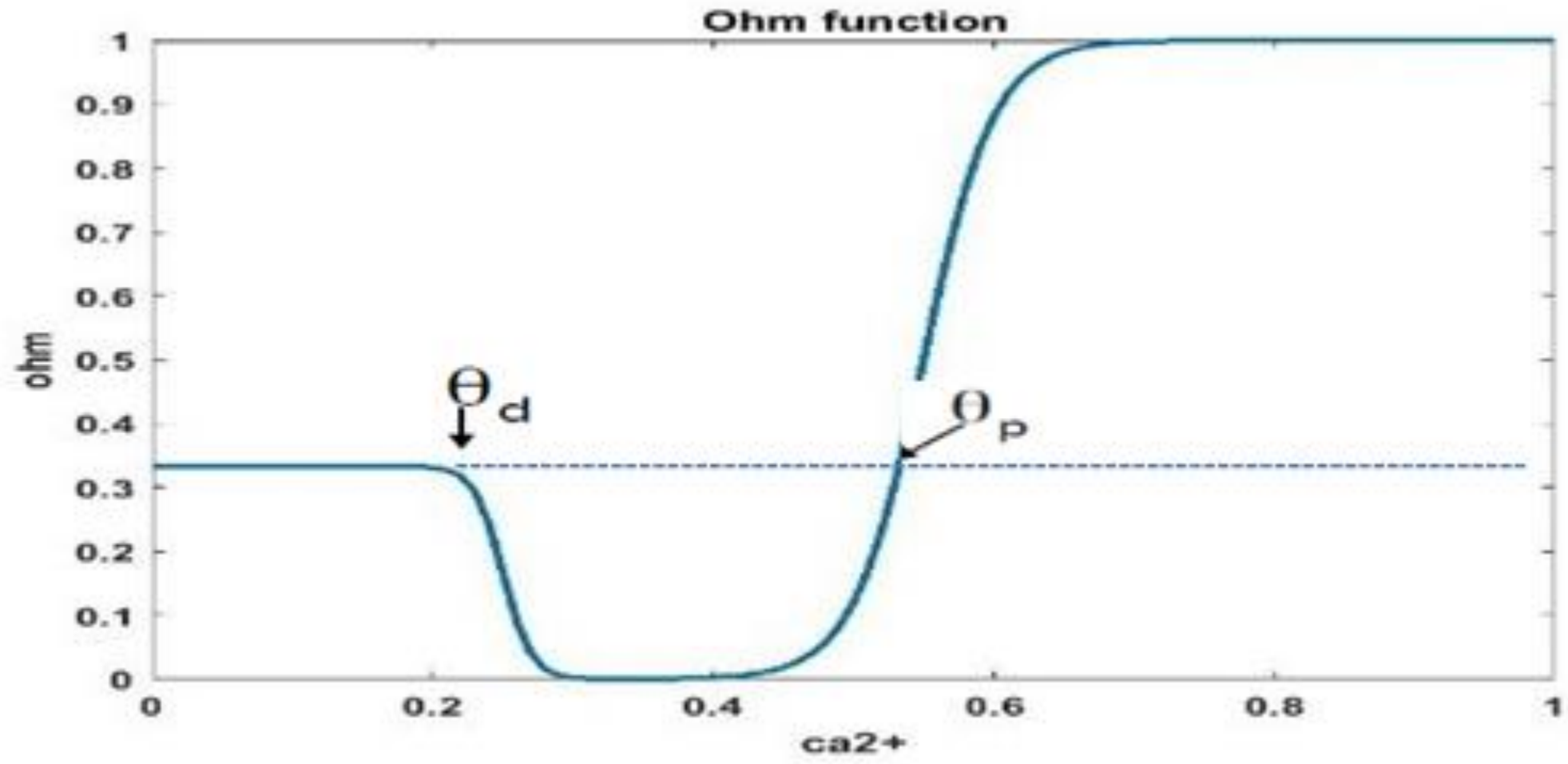

**Fig 2.1. The ohm function (Top) in the synaptic weight modification equation**

.

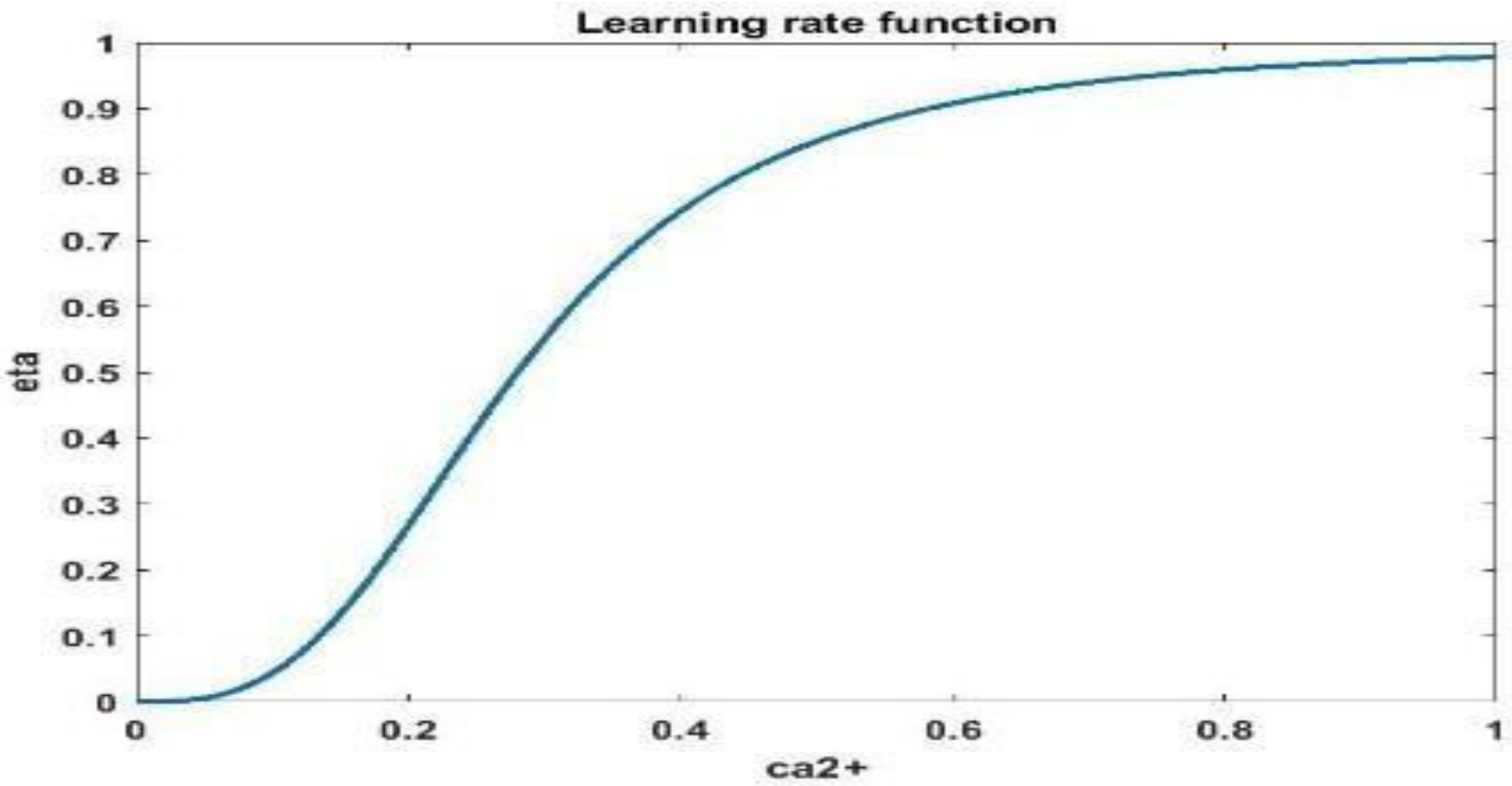

**Fig 2.2. The learning rate function in the synaptic weight modification equation.**

The weights are manipulated based on spike timing:

Case 1: When tpre ≤ tpost within set window: ( $[Ca] > \theta_p$ ) Synaptic weight is strengthened by $dW_i$.

Case 2: When tpre > tpost within set window: ( $\theta_d < [Ca] < \theta_p$ ) Synaptic weight is depressed by $dW_i$

Case 3: When tpre and tpost are not within set window: Synaptic weight is unchangedtpre ➔ pre-synaptic spike timetpost ➔ post-synaptic spike time

## 2.6 Peripheral Architecture

The proposed algorithm is simulated with an assumption of programmable hardware, the details of which are listed below:

1. Array of blue light lasers

   A cross section of lasers with each pinpointed at one neuron. Strength: $1mW/mm^2$

   Time: 5ms pulse

   Type: Blue light laser 473 nm

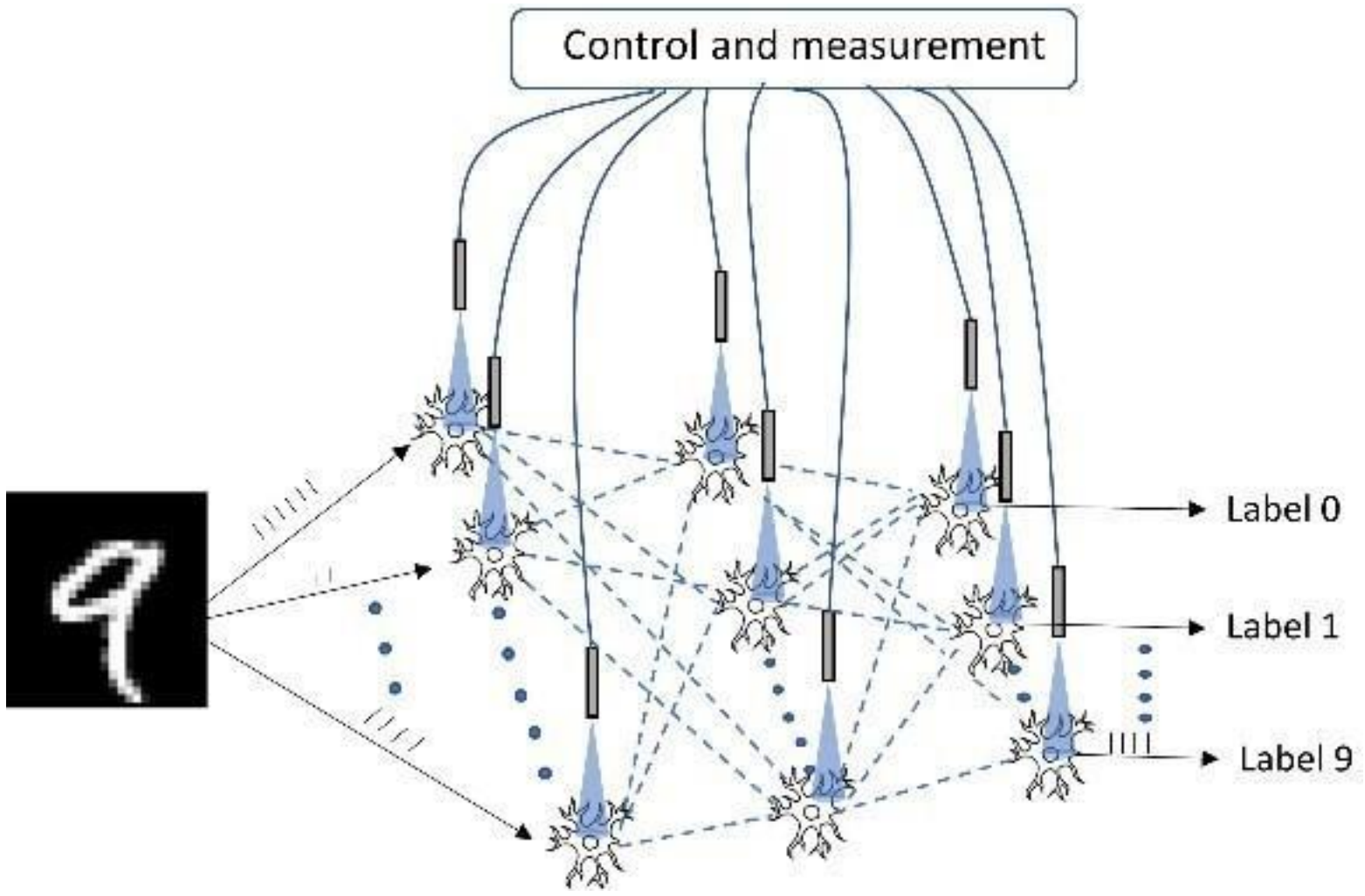

**Fig 2.3. Proposed architecture for the training of the neural network. A three-layer neural network, with pixel intensities converted to spike trains for input. The indirect training and readout is done by an array of LEDs and various recording hardware.**

Hardware developed by groups like [22][24] have been shown to stimulate with single neuron precision, which is a requirement for the algorithm proposed.

2. Synaptic and membrane voltage readout:

   The setup includes readout hardware such as intracellular electrodes, patch clamps and fluorescence imaging systems to measure membrane voltages and synaptic strength for each iteration. Hardware developed by groups like [21][26] have been shown to record individual synaptic strength over thousands of neurons.

# Chapter 3: Algorithm

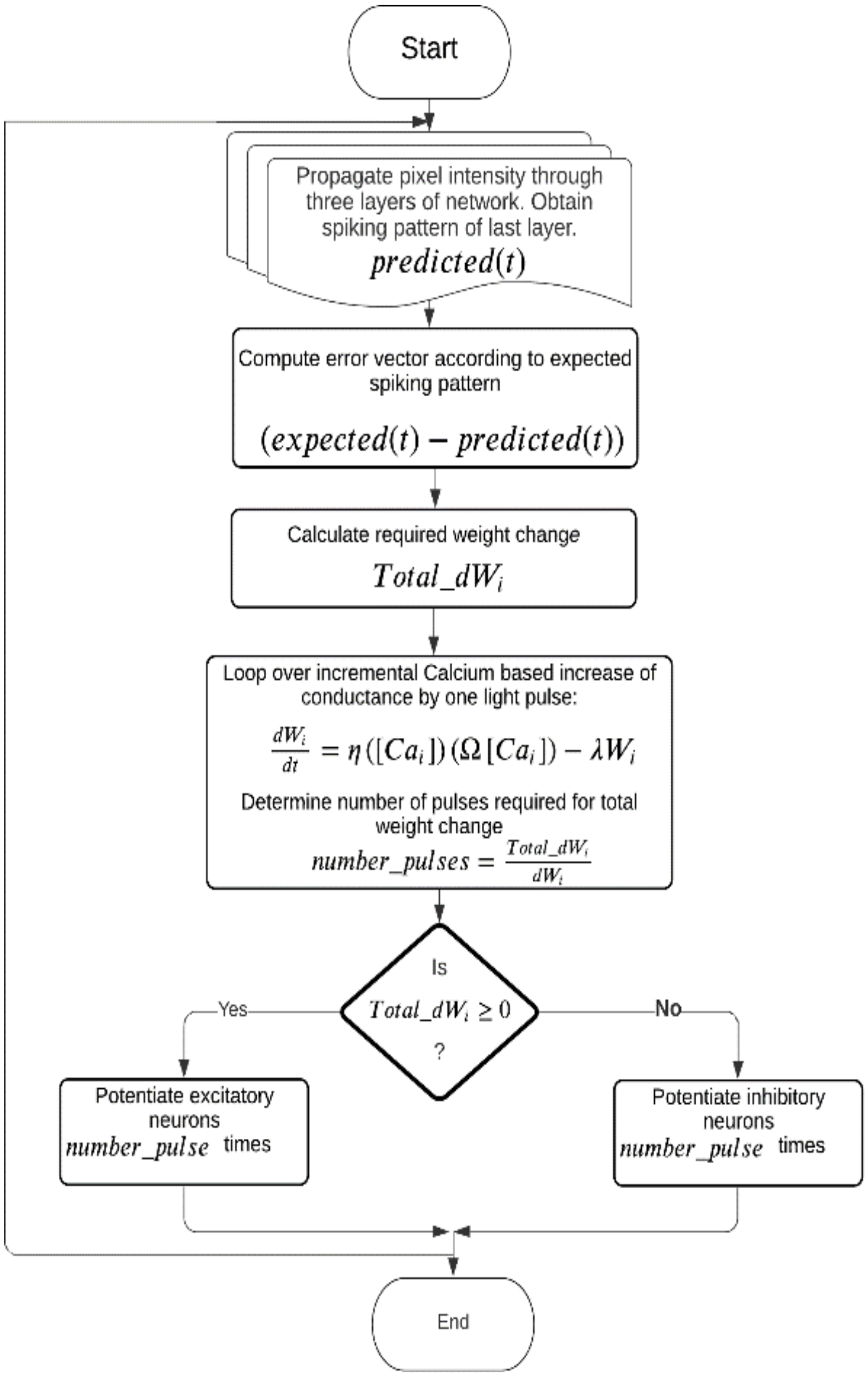

**Fig 3.1 Basic Flowchart of training algorithm**

A forward pass is conducted with an electrical input, over three layers of the neural network. The error is calculated based on the expected results and the predicted results. Owing to the all-or-nothing nature of biological neurons, the outer layer error will be determined by checking if the expected neuron has spiked or not. This error is then back propagated, and the hidden layer error is calculated. Once error is calculated, the required change in conductance is determined. If the change in conductance determined is negative, this implies that the synaptic input to the postsynaptic neuron must be reduced. This can be achieved by strengthening the conductance of the inhibitory neuron with respect to the postsynaptic neuron. Conversely, if the change in conductance determined is positive, this implies that the synaptic input to the postsynaptic neuron must be increased, which can be achieved by strengthening conductance of the excitatory neuron with respect to the postsynaptic neuron. The total conductance change is achieved by exciting the presynaptic and postsynaptic neurons together with a calculated number of light pulses within the potentiation window based on STDP principles. The whole process is repeated for a predetermined time period and for all training samples, to find a global optimum. The equations used in the algorithm are further detailed in the next section.

## 3.1 Algorithm

In this section we elaborate on the algorithm proposed to indirectly train live neural networks. As discussed earlier, the Algorithm consists of the forward pass, the backward pass and the indirect training method

Each epoch-Sample 50 points randomly from dataset

*Mean squared err*
$$\mathrm{MSE} = \frac{1}{N}\sum_{k=1}^{N}\left(\frac{1}{T}\sum_{t=1}^{T}\xi^{k}(t)\right)^{2}, \quad \xi^{k}(t) = \sum_{i}\xi_{i}^{k}(t)$$

*Iterate over t=1:dt:T*

***** Forward Pass *********

*------Layer 1: (input layer):-*

$$membrane\ potential(t) = \mathrm{Neuron}(pixel\ intensity)$$

$$spikes_inp(t) = \begin{cases} 0 & membrane\ potential\ input(t) \leq threshold_{inp} \\ 1 & membrane\ potential\ input(t) > threshold_{inp} \end{cases}$$

*Calculate synaptic input for Layer2:*

$$synap_Total_Input(t) = synap_InpHidden_excitatory(t) - synap_InpHidden_inhibitory(t)$$

*------Layer 2: (hidden layer):-*

$$membranePotentialHid(t) = Neuron(synap_Total_Input(t))$$

$$spikes_hid(t) = \begin{cases} 0 & membrane_potentialHidden(t) \leq threshold_hid \\ 1 & membrane_potentialHidden(t) > threshold_hid \end{cases}$$

*Calculate synaptic input for Layer:*

$$synap_Total_Hidden(t) = synap_HiddenOutput_excitatory(t) - synap_HiddenOutput_inhibitory(t)$$

*------Layer 3: (Output layer):-*

$$MembranePotentialOutput(t) = \mathrm{Neuron}(synap_Total_Hidden(t))$$

$$spikes_out(t) = \begin{cases} 0 & membrane_potentialoutput(t) \leq threshold_out \\ 1 & membrane_potentialoutput(t) > threshold_out \end{cases}$$

******** Backward Pass computations ***********

*Calculate required change in synaptic conductance:*

$Predicted_output(t) = spikes_out(t)$

$Error_op(t) = Expected_output(t) - Predicted_ouptut(t)$

$Error_hid = \sum(g_HO_exc - g_HO_inh) * Error_op(t)$ *for all hidden neuron that spiked*

$\Delta gHO = \sum spikes_hid\ (t - \varepsilon: t) * Error_op * \ \mu$

$\Delta gIH = \sum spikes_inp\ (t - \varepsilon: t) * Error_hid * \mu$

If $\Delta gHO > 0 : \Delta wHO_excite = \Delta gHO$

$\Delta gHO < 0 : \Delta wHO_inh = \Delta gHO$

If $\Delta gIH > 0 : \Delta gIH_excite = \ \Delta gIH$

$\Delta gIH < 0 : \Delta gIH_inh = \ \Delta gIH$

*Where* $\varepsilon$ *is the STDP window and* $\mu$ *is the learning rate.*

******** Backward Pass ***********

$$\frac{dW_{i_single}}{dt} = \eta([Ca]_i)(\Omega([Ca]_i) - \lambda W_i$$

$factorID\ (\Delta gIH, \Delta gHO) = \Delta\ TotalConductance\ (\Delta gIH, \Delta gHO)/dW_{i_single}$

*Iterate over t=1: max(*$factorID_HO$*)*

*Optically activate neuron pairs with calculated number of singular light pulses(*$factorID$*)to achieve required conductance changed*

## Chapter 4: Results

### 4.1 Optogenetically active neuron:

The individual neuron model with embedded CHR2 channels was excited with simulated light pulses of intensity 1mW/mm$^2$, 5mS pulse width and a period of 55mS. It produces single spikes with low rise time, comparable to spikes produced by electrical pulses. This precise strength and pulse width was utilized for training in all the training examples.

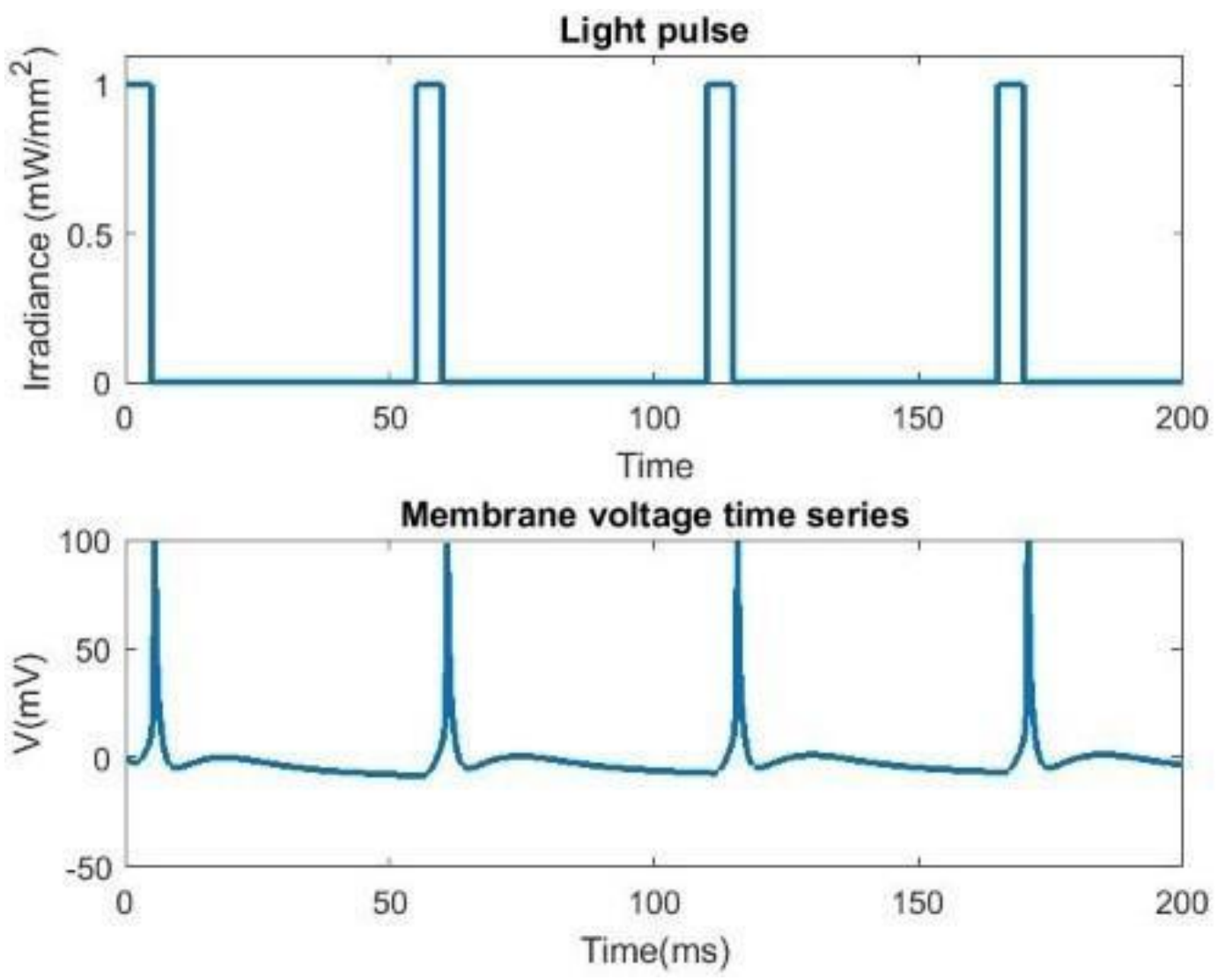

**Fig 4.1. Simulation of single neuron stimulated with 5mS blue light pulse of strength 1mW/mm$^2$ to produce single spike**

The algorithm was tested over two different datasets: XOR and MNIST to test accuracy.

### 4.2 XOR dataset:

The network consists of 2 excitatory input neurons, 2 inhibitory input neurons, and 2 excitatory output neurons. Dataset used:

{(0.2,0.2), (0.2,1.0), (1.0,0.2), (1.0,1.0)}; Expected response: {0,1,1,0}. The value of 0.2 was used instead of 0, to provide a mild excitation. The number of hidden neurons was varied {2,5,10,20,40,80,100} with an equal number of excitatory and inhibitory neurons. The accuracy was found to saturate for hidden neurons greater than or equal to 20 excitatory and 20 inhibitory neurons. The Learning rate was varied over a range of values with the number of hidden neurons fixed at 20 excitatory and 20 inhibitory neurons. The network was trained for 150 epochs, with a 50mS period for each input. The results are shown below.

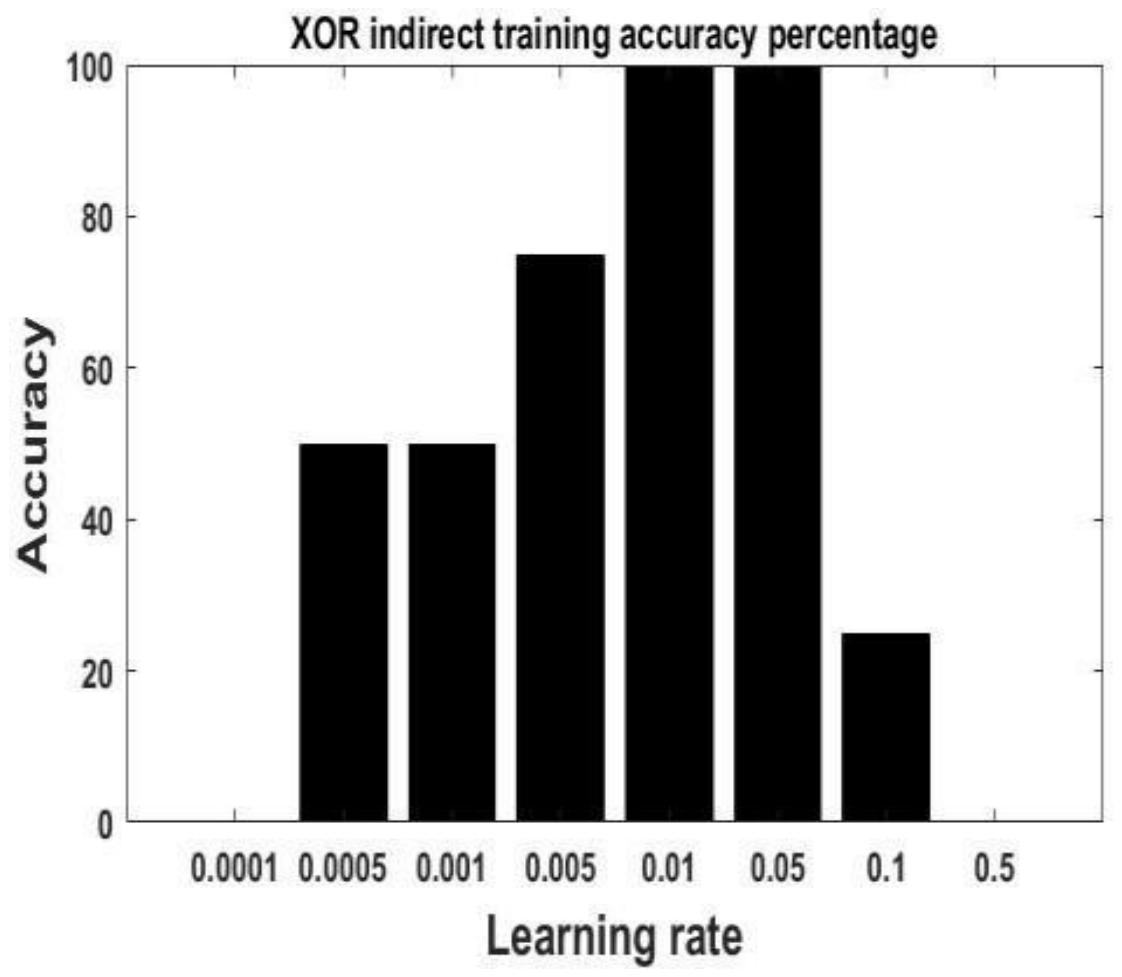

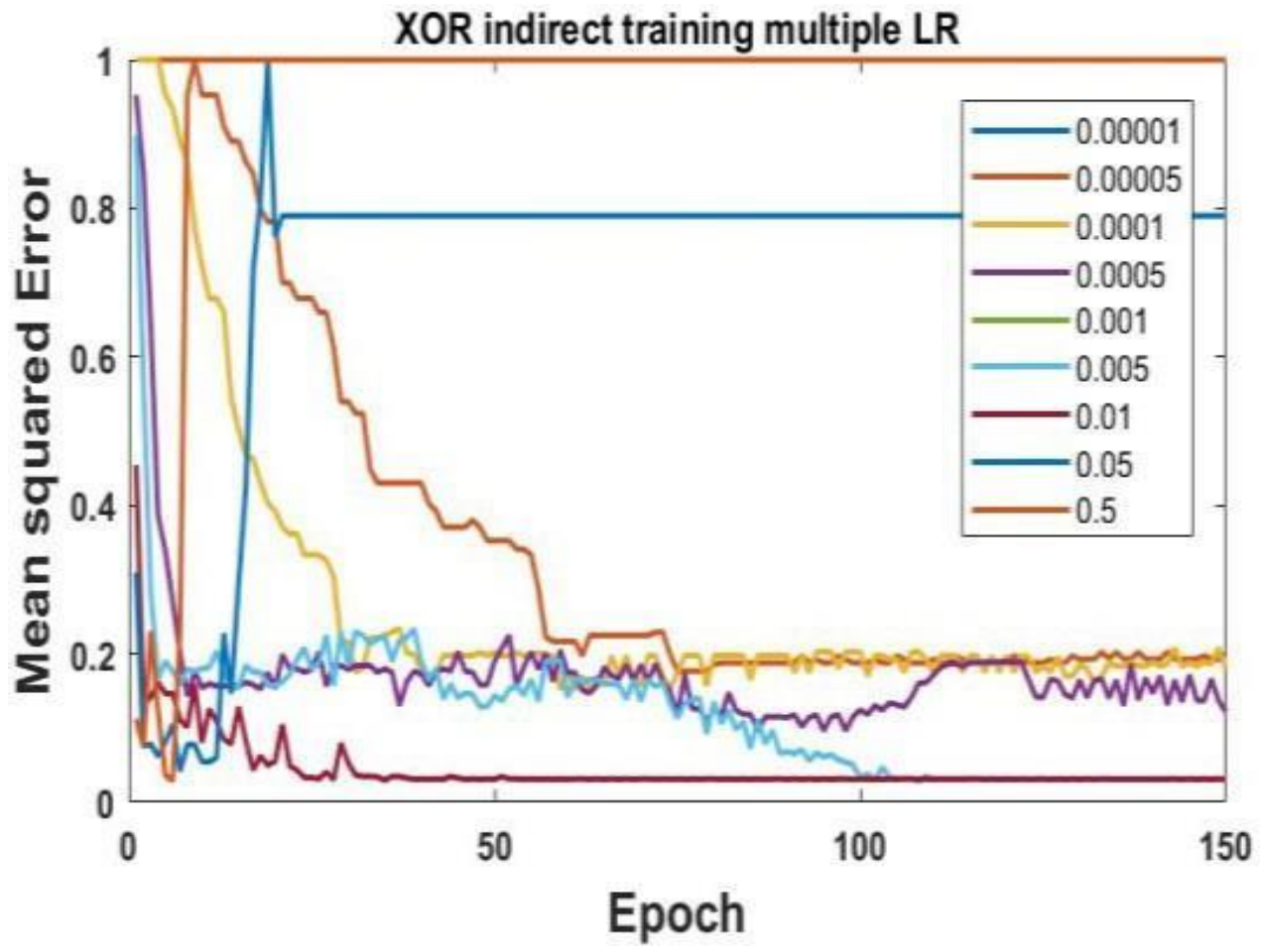

**(a)** **(b)**

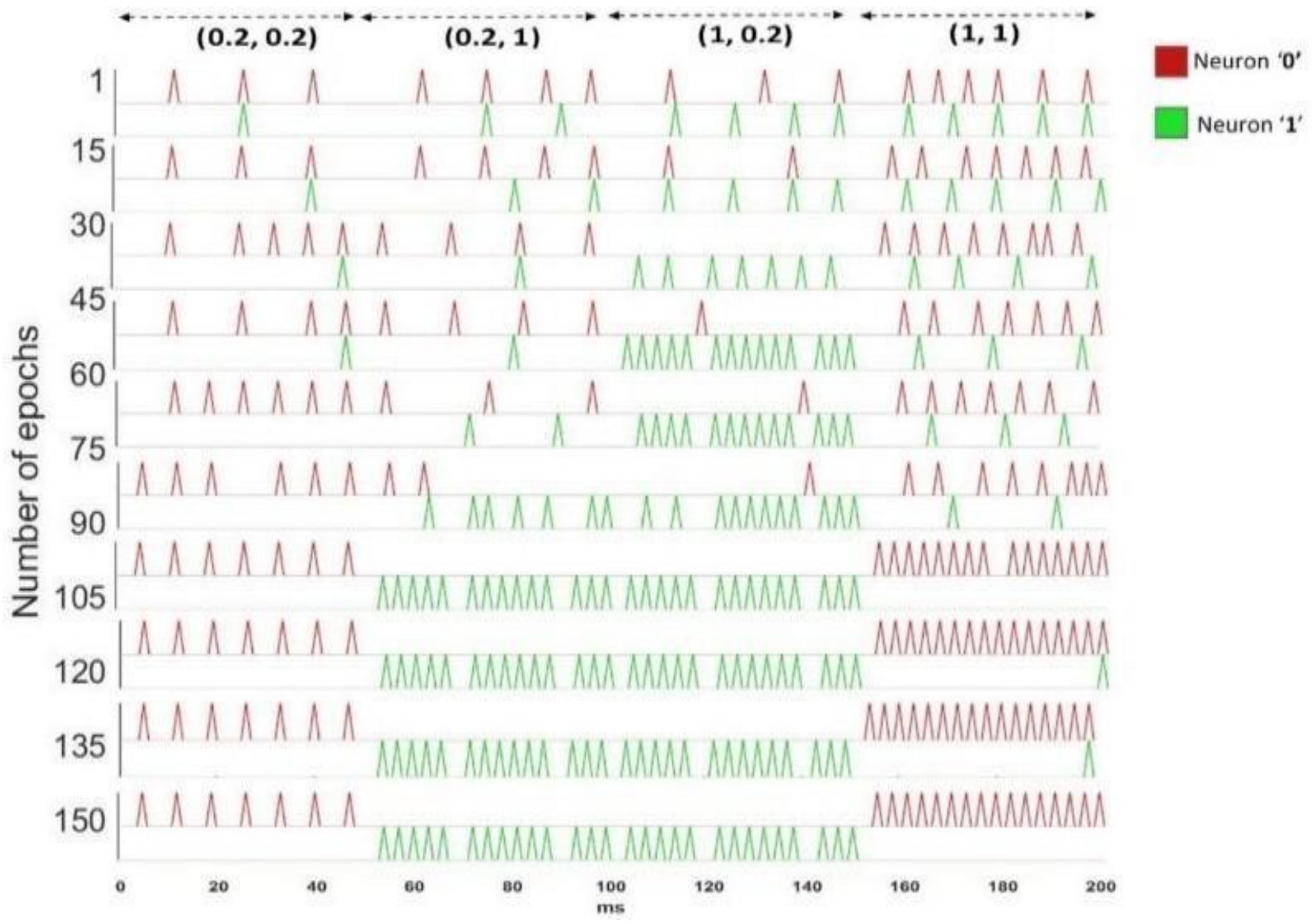

**(c)**

**Fig 4.2. (a). Accuracy in percentage vs Learning rates (b) Mean squared error vs Learning rates (c) Spiking of Neuron '0' and Neuron '1' over epochs 0 to 150 for fixed learning rate of 0.01 for XOR dataset**

The inputs were spike trains of electric signals with amplitudes proportional to {0.2, 1}. The process, as described in *Algorithm*, was executed in two phases: A forward pass to observe predicted results from the optogenetic neural network; A backward pass to train the network with pulses of blue light. A maximum accuracy of 98% was obtained. The fastest convergence and best accuracies were noted for learning rates 0.05 and 0.01. At the end of training, the expected neuron is observed to have the highest spiking rate while all other output neurons are silent. As seen in fig 5(c), we observe the erratic spiking of output neurons at epoch 1, which is trained to obtain the required output spiking pattern by epoch 150.

### 4.3 MNIST dataset:

The network consists of 784 excitatory input neurons, 784 inhibitory input neurons, and 10 output neurons. Dataset used: MNIST; Expected response: {0,1,2,3,4,5,6,7,8,9}. The network is trained on 60k data points and tested on 10k data points over 100 epochs, with a training set size of 50 samples per epoch, and a 50mS period for each input. The number of hidden neurons was varied 10 through 2000 with an equal number of excitatory and inhibitory neurons. The accuracy was found to increase linearly and saturate for hidden neurons greater than or equal to around 300 excitatory and 300 inhibitory neurons (shown in Fig 4.3a). The fastest convergence was noted for total hidden neurons greater than or equal to 1000 (500 inhibitory and 500 excitatory).

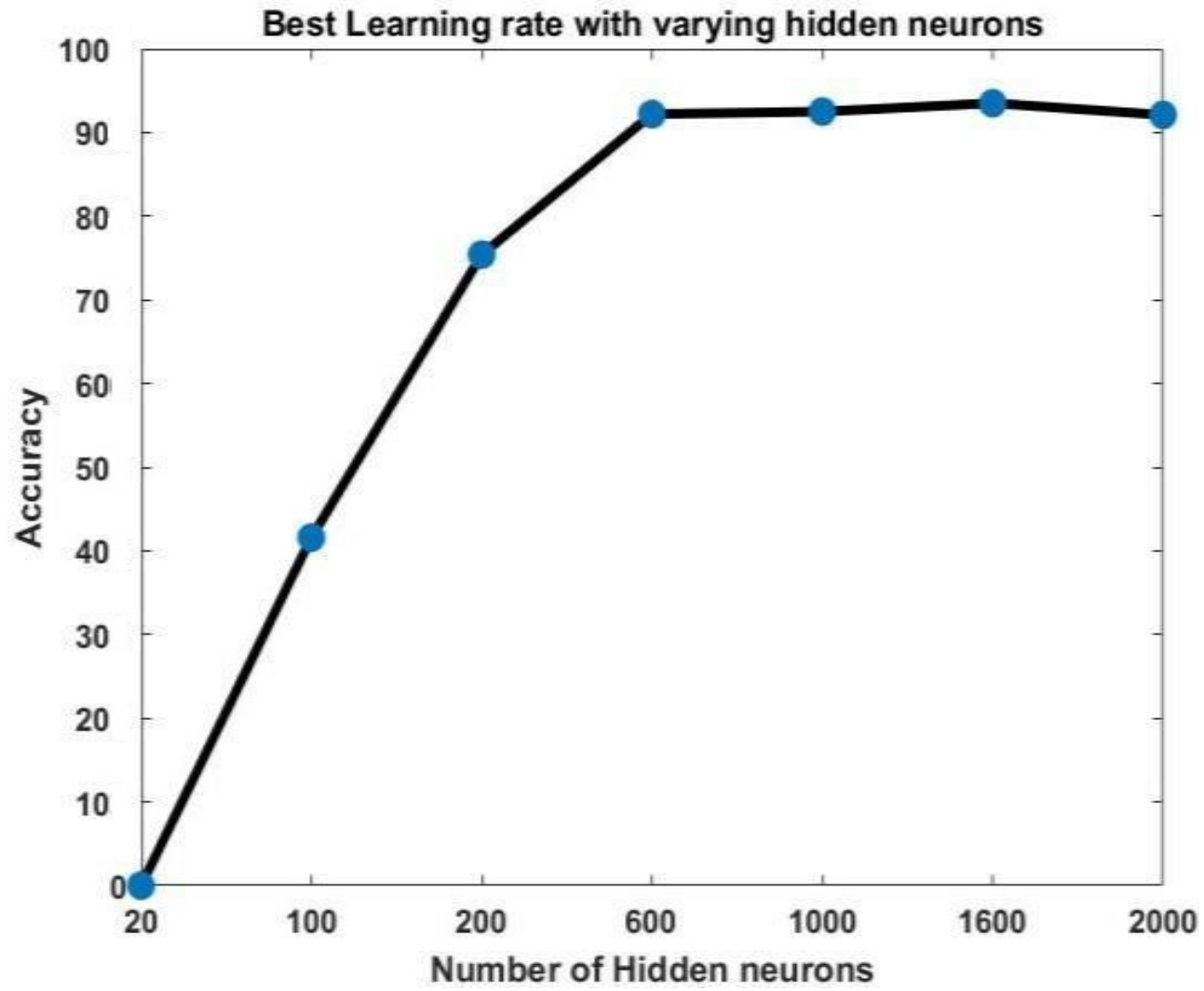

(a)

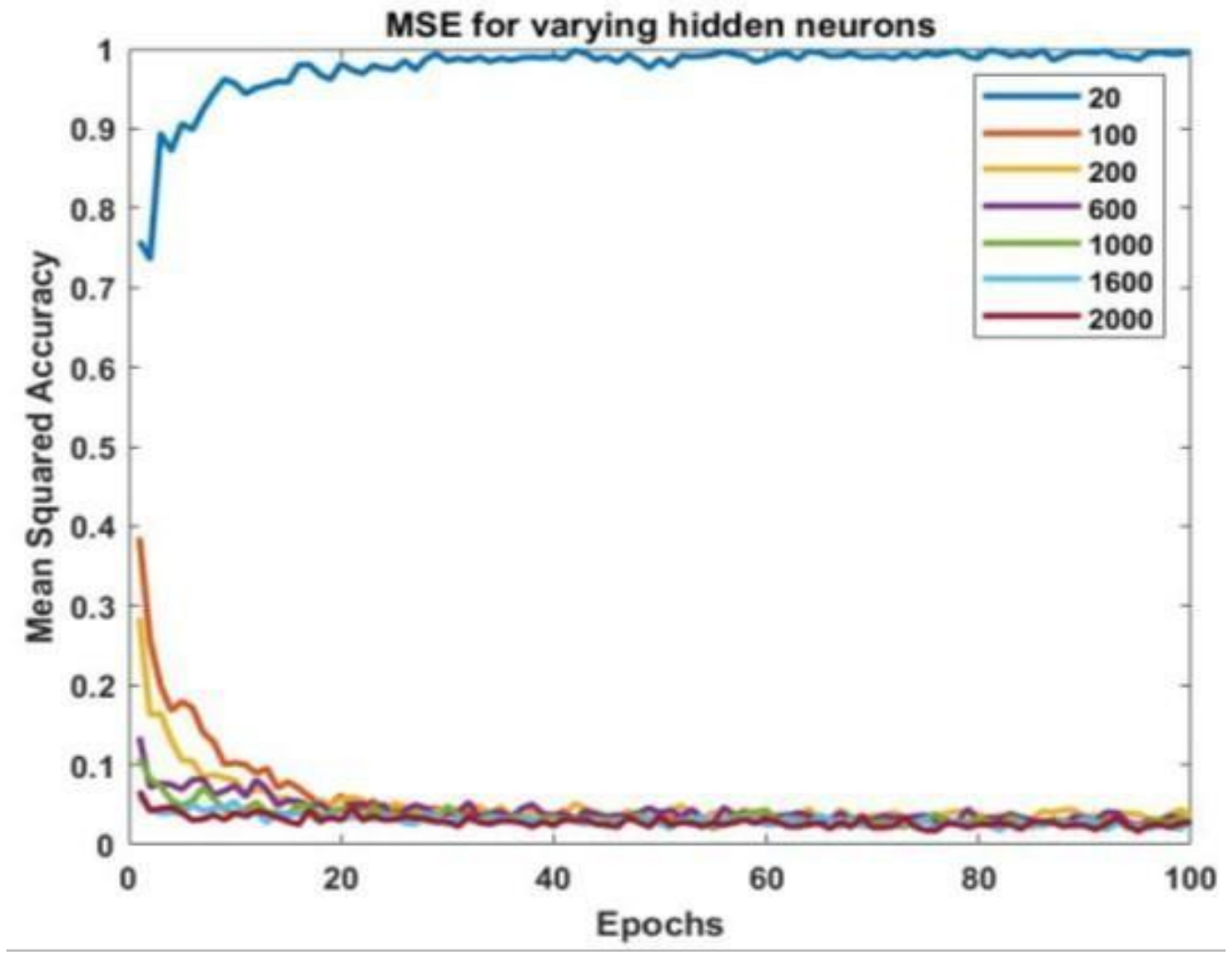

(b)

**Fig 4.3. (a). Accuracy in percentage vs number of hidden neurons (b) Mean squared error vs number of hidden neurons for complete MNIST dataset.**

The Learning rate was varied over a range of values {1.0e-6 …. 5.0e-3} with the number of hidden neurons fixed at 500 excitatory and 500 inhibitory neurons. The results are shown below.

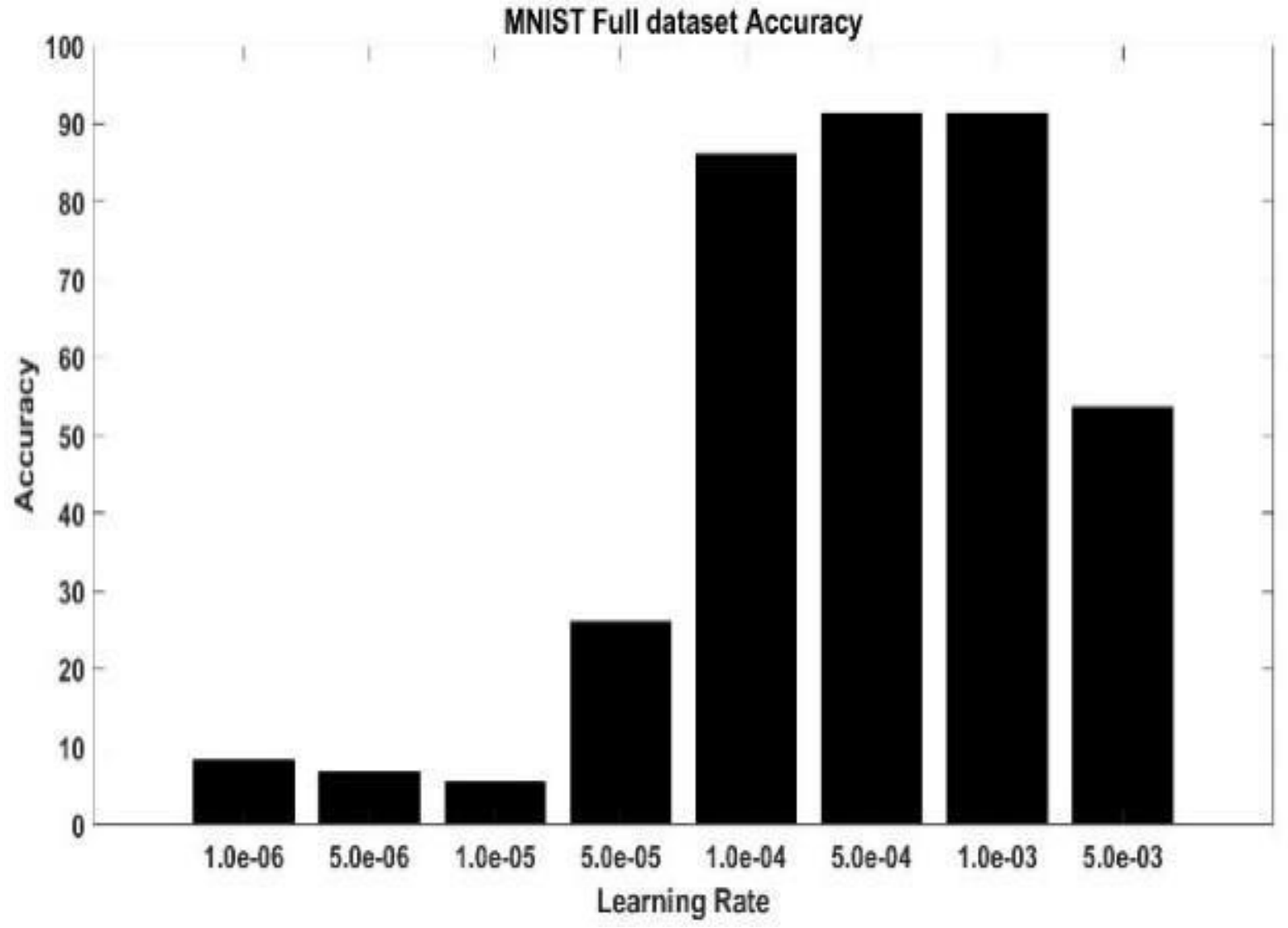

**Fig 4.4.(a). Accuracy in percentage vs learning rates**

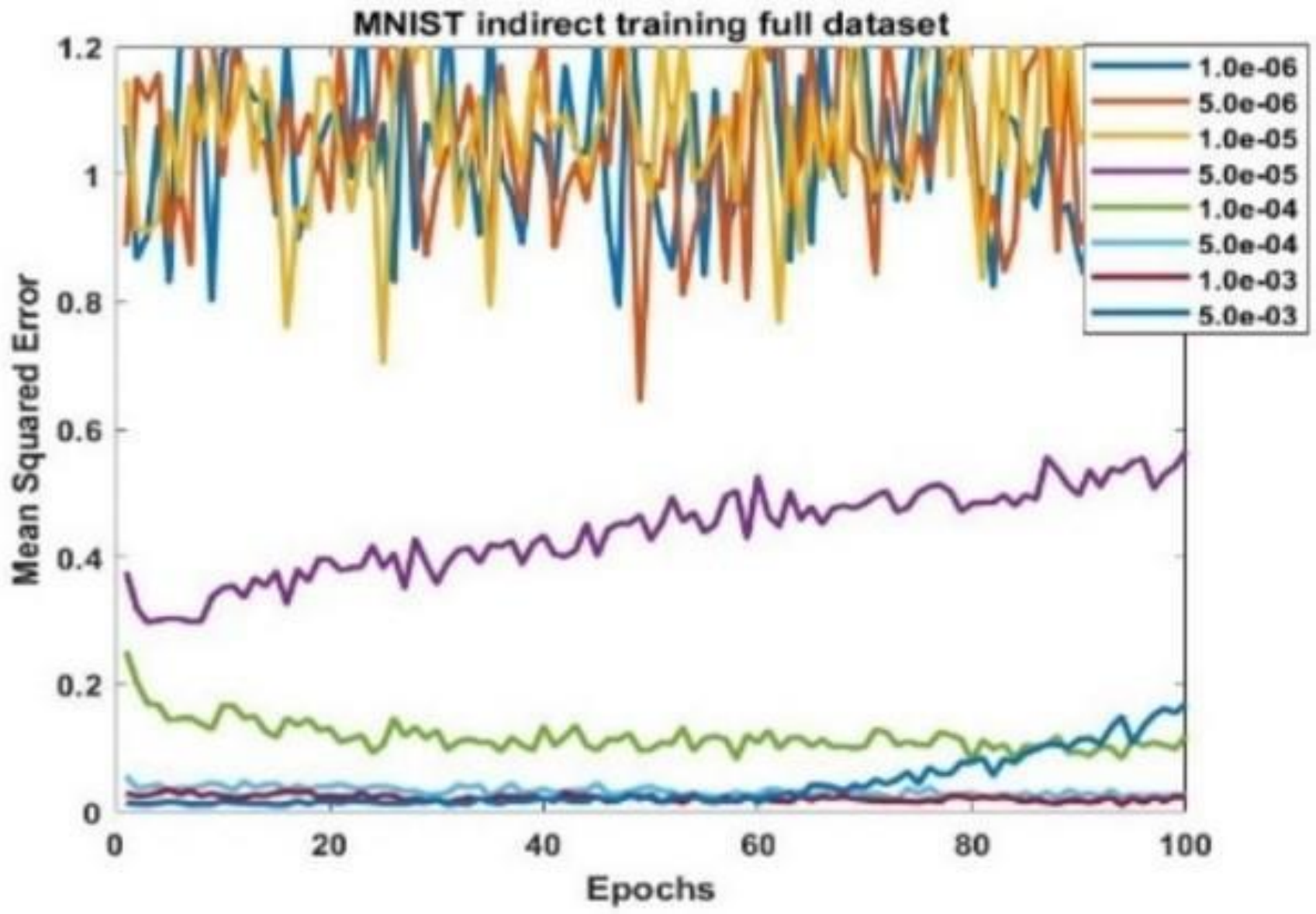

**Fig 4.4. (b) Mean squared error vs Learning rates for complete MNIST dataset**

The inputs were spike trains of electric signals with amplitudes proportional to pixel intensities. The pixel intensities were normalized in range [0 1]. The process, as described in *Algorithm,* was executed in two phases: A forward pass to observe predicted results from the optogenetic neural network; A backward pass to train the network with pulses of blue light. A maximum accuracy of 92% was obtained. The fastest convergence and best accuracies were noted for learning rates 0.0005 and 0.001. At the end of training, the expected neuron is observed to have the highest spiking rate while all other output neurons are silent. In fig 4.6(a), 4.6(b), 4.6(c),4.6 (d), we observe the output obtained for MNIST images of 1, 8, 3 and 7. The expected neuron is observed to have the largest number of spikes in the 50mS period while all the other output neurons are dormant and have not reached the required spiking threshold.

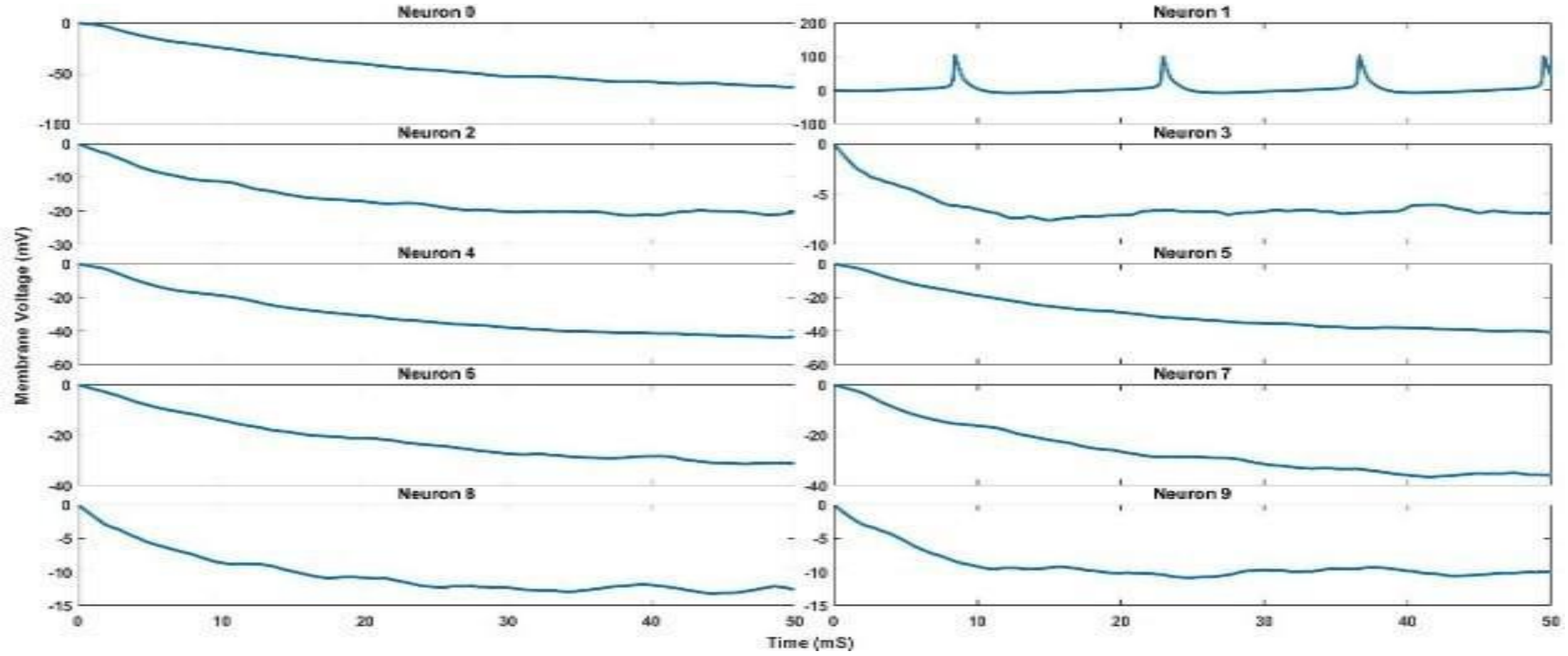

**Fig 4.5. (a). Spiking patterns of output neurons for inputs of (a)**

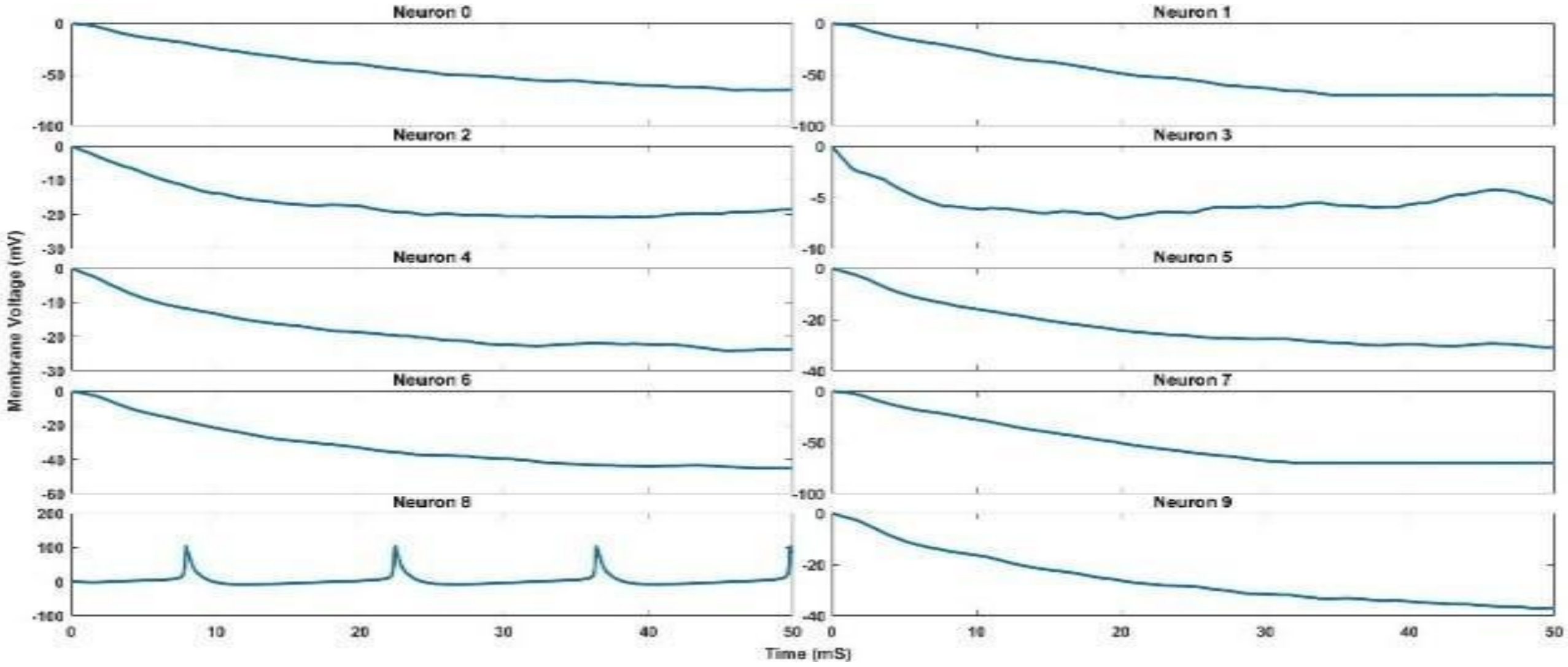

**Fig 4.6 (a)** **Fig 4.6 (b)**

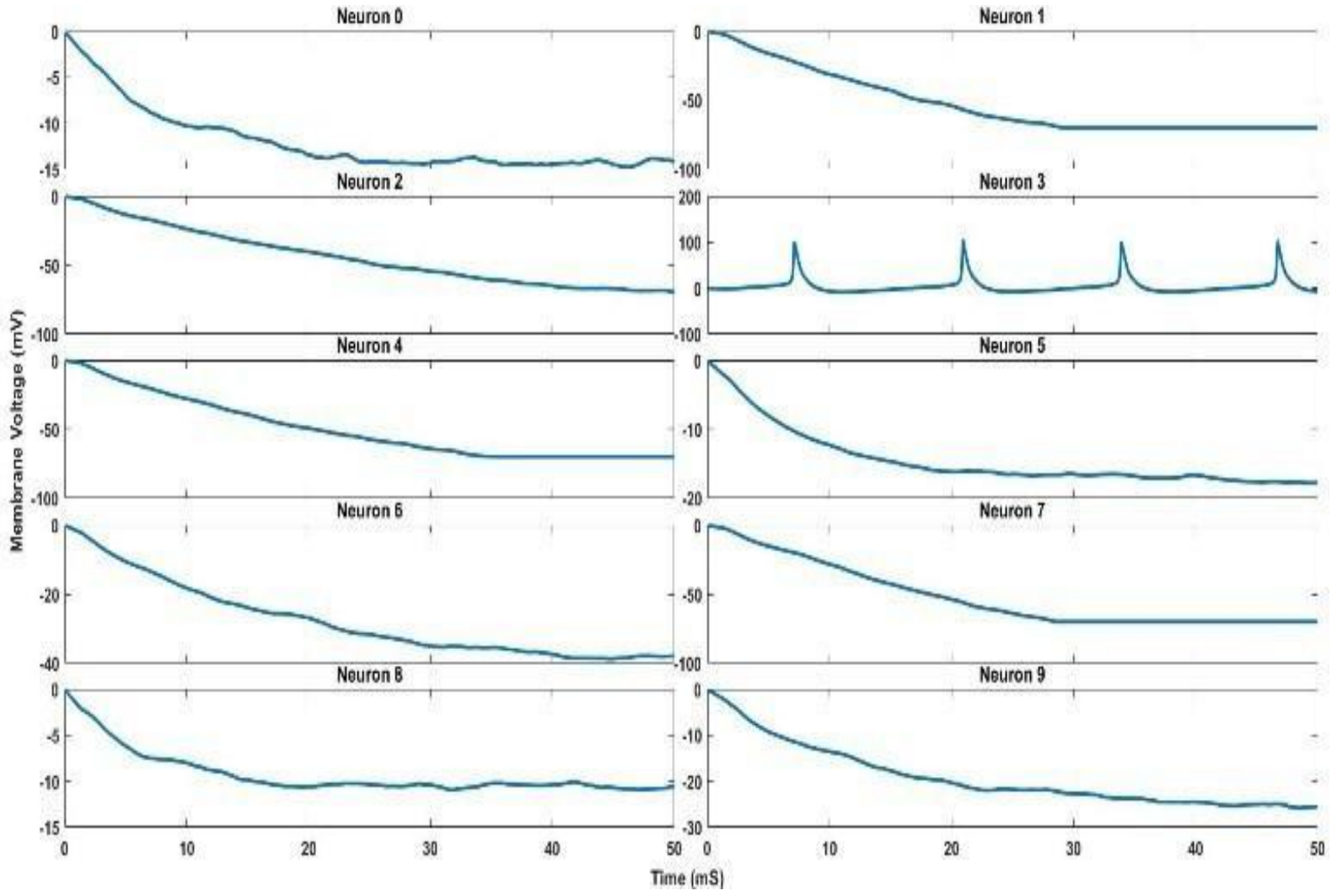

**Fig 4.6 (c)**

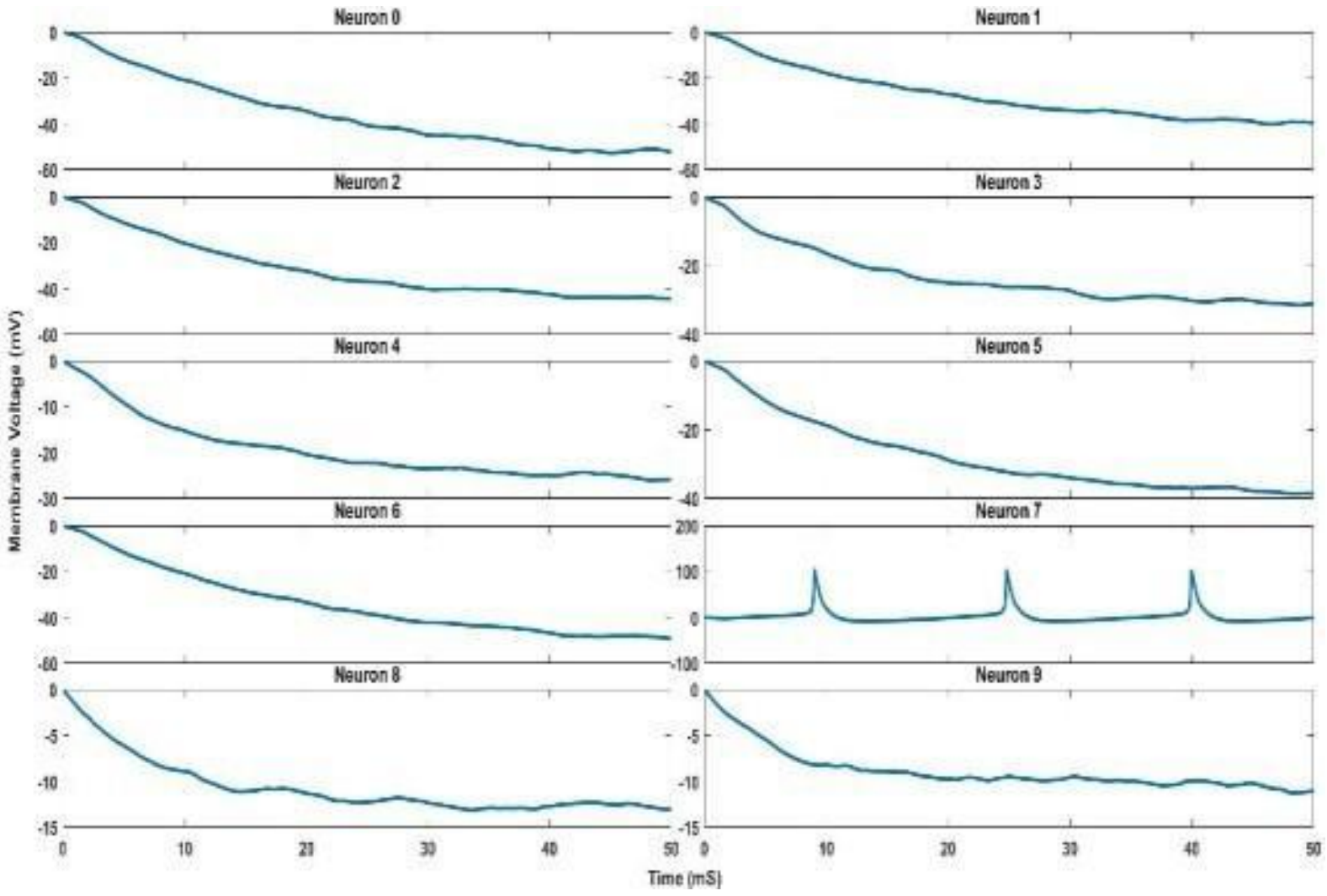

**Fig 4.6 (d)**

**Fig 4.6(a). MNIST input '1' (b) MNIST input '8' (c) MNIST input '3' (d) MNIST input '7'**

Additionally, a compressed dataset was used as input to reduce the number of input neurons required in the network. The original image matrix of 28 × 28-pixel values is reduced to a 14 × 14 matrix by Max pooling. This reduces the required input neurons to 196 inhibitory neurons and 196 excitatory neurons. The results are as shown below.

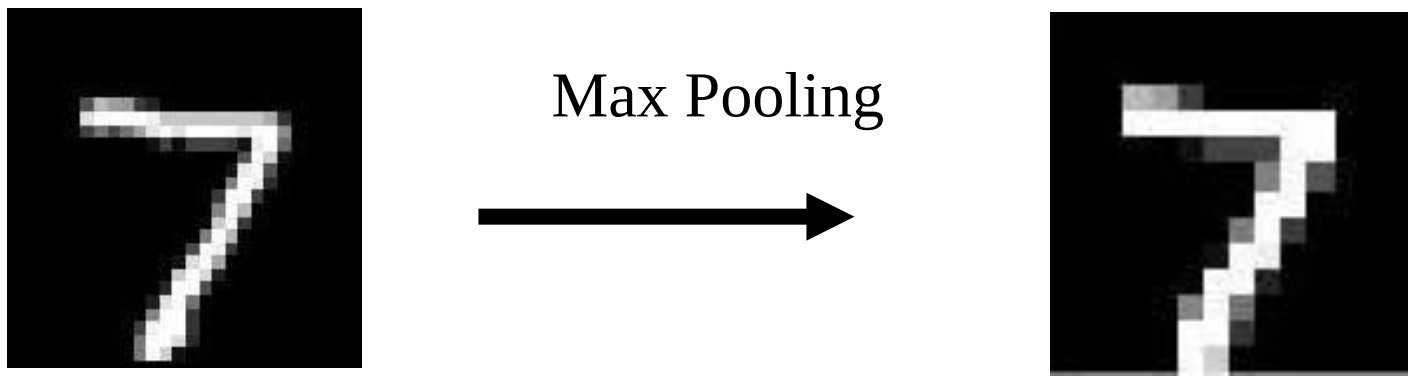

**Fig 4.7. Compression of 28*28 matrix of pixel intensities to 14*14 matrix of pixel intensities using max pooling**

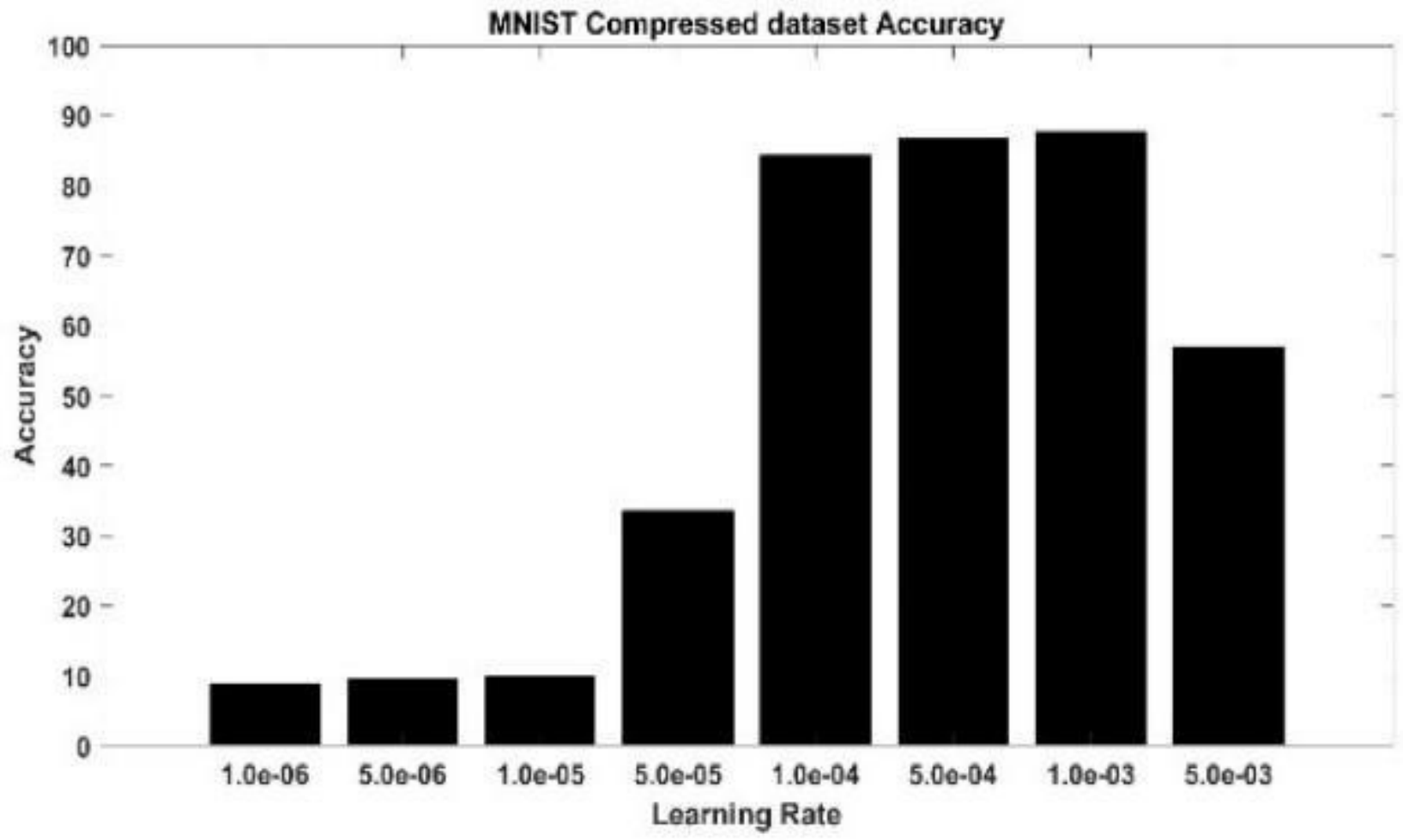

**Fig 4.8 (a). Accuracy percentage vs learning rates**

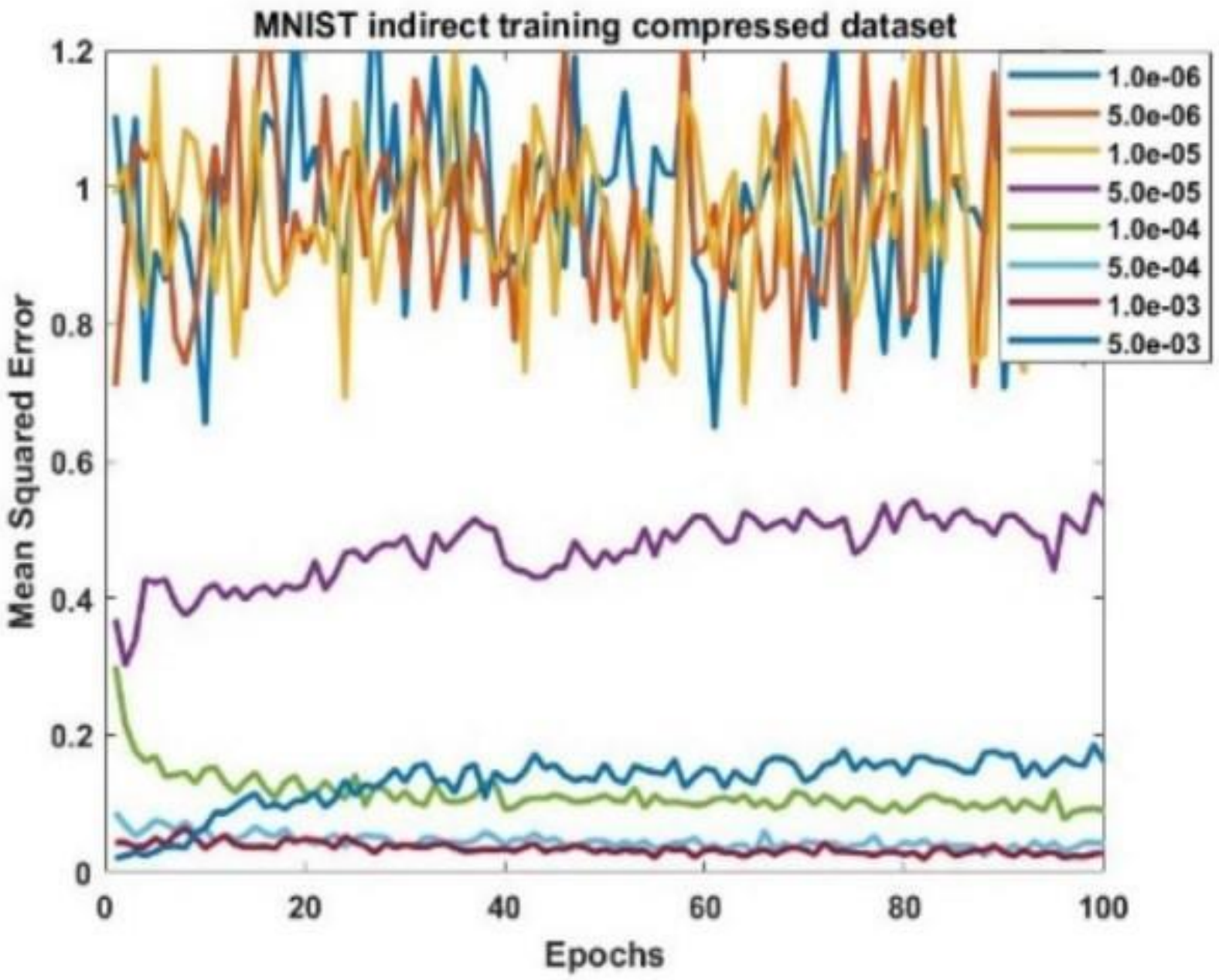

**Fig 4.8 (b) Mean squared error vs Learning rates for compressed MNIST dataset.**

The trends observed are very similar to data obtained with a full image dataset. The fastest convergence and best accuracies are observed again for learning rates 0.0005 and 0.001. The best accuracy obtained was slightly reduced at an average high of 88%.

# Chapter 5: Conclusion and Future Work

The proposed STDP based indirect training with backpropagation was found to produce results comparable to conventional Artificial neural networks. Overall, a high accuracy of 92% was obtained on the full MNIST dataset using ~3000 neurons. The proposed algorithm addresses the potential implementation in biological neural networks by including three specific methods: indirect training, restriction to only positive conductance and implementation of optogenetic control. With indirect training, the model takes into account the spatio-temporal nature of cell firings and the inability of direct manipulation of neuron conductance in its learning rule and proposes a manipulation of STDP rules to produce the required synaptic weight changes. By artificially spiking a presynaptic and postsynaptic neuron within a certain window, we effectively employ nature's own learning methods to train our neural network. A calcium-based learning rule provides a better projection of required number of training pulses and also aids in readout of synaptic strength by possible calcium imaging. In traditional Artificial neural networks, constraints of positive only conductance, limits the decision surfaces that the neural network can learn in and results in lower accuracy. In ANNs, this limitation has been partially addressed by pruning, regularization, varied activation functions and other methods of computation that cannot be implemented in live neural networks. In our work, this constraint is addressed by inclusion of inhibitory neurons. By strengthening conductance of inhibitory neurons, we pass a larger negative input to the postsynaptic neuron, effectively hyperpolarizing it and lowering chances of spiking. The inclusion of optogenetic activation provides the high-level precision required for single neuron excitation, while leaving surrounding neurons unperturbed as opposed to interference caused by electric stimulation. Past works have worked on direct manipulation of SNNs [27][28][29], direct training using STDP [14], Single layer training of biologically plausible neural networks[30] and basic motor control algorithms[9][11]. Expanding on those works, the method proposed in this paper has an increased accuracy, is developed with consideration of biological constraints and is capable of identifying complex images.

To explore the potential of live neural networks in their computational capabilities and energy efficiency, a biological plausible multi-layer neural network training algorithm is proposed in this work. The neuron model, synapse model and methods of training have been kept as biologically accurate as possible in simulations. Optogenetic methods are included for precise spatio-temporal activation. Its obtained accuracy of 92% on the MNIST dataset is comparable to accuracy produced by ANNs. Possible applications of this algorithm can be extended to hybrid robots, live computational chips and can potentially be extended to neuroprosthetics and treatment of neurodegenerative diseases.